\documentstyle[12pt,colordvi,epsfig,axodraw]{article} 
\let\jnfont=\rm
\def\NPB#1,{{\jnfont Nucl.\ Phys.\ B }{\bf #1},}
\def\PLB#1,{{\jnfont Phys.\ Lett.\ B }{\bf #1},}
\def\EPJC#1,{{\jnfont Euro.\ Phys.\ J.\ C }{\bf #1},}
\def\PRD#1,{{\jnfont Phys.\ Rev.\ D }{\bf #1},}
\def\PRL#1,{{\jnfont Phys.\ Rev.\ Lett.\ }{\bf #1},}
\def\MPLA#1,{{\jnfont Mod.\ Phys.\ Lett.\ A }{\bf #1},}
\def\JPG#1,{{\jnfont J.\ Phys.\ G}{\bf #1},}
\begin{document}

\begin{center}

\Blue{\Large \bf Probing new physics from top quark FCNC}

\Blue{\Large \bf processes at linear colliders : a mini review
\footnote{\Green{Talk given at APPI2004, Feb 16-19, Iweta, Japan}}
}
\vspace*{1.0cm}
\normalsize

Jin Min Yang
\vspace*{0.50cm}

{\it Institute of Theoretical Physics, Academia Sinica, Beijing, China;} \\
{\it Department of Physics, Tohoku University, Sendai, Japan}
\end{center}
\vspace*{1.0cm}

\normalsize
\begin{center} \Blue{\it{Abstract}} \end{center}
\small

\it{ We briefly review the studies on the top quark FCNC 
processes at a next-generation linear collider. Such processes, including various FCNC top quark
rare decays and top-charm associated productions, are extremely suppressed in the 
standard model (SM) but could be significantly enhanced in some extensions. 
We compared the predictions from different typical new physics models: the SM,  
the minimal supersymmetric model (MSSM), the general two-higgs-doublet model
(2HDM) and the topcolor-assisted technicolor (TC2) model.
Our conclusion is: \Blue{(1)} While all the new physics models can enhance the rates 
by several orders relative to the SM predictions, the TC2 model predicts much 
larger rates than other models; \Blue{(2)} The optimal channel for probing the top quark 
FCNC is the top-charm associated production in $\gamma \gamma$ collision, which occurs 
at a much higher rate than $e^+ e^-$ or $e^- \gamma$ collision and can reach the detectable 
level for a large part of the parameter space. }
\normalsize

\Blue{\section{Introduction}}

~~~~A next-generation linear collider (LC) will be an ideal machine for precisely testing the SM and 
probing new physics.  In such a collider, in addition to $e^+ e^-$ collision, we can also realize 
$\gamma \gamma$ collision and $e^- \gamma$ collision with the photon beams generated by 
the backward Compton scattering of incident electron- and laser-beams \Blue{\cite{JLC}}. 

Probing new physics is the most important goal for an LC. 
There are two ways for probing new physics: One is the direct search for new particles; 
the other is the indirect search from the quantum effects caused by new particles
in some SM processes.  Of course, if the new particles 
are beyond the  energy threshold of the collider, the only possible 
way to reveal their existence is through the indirect search from their quantum effects.
  
Top quark may serve as a window for probing new physics. 
On the experimental side, since the measurements of top quark properties 
at the Fermilab Tevatron have small statistics, there remains plenty of room for 
new physics in the top quark sector. On the theoretical side, it is reasonable 
to expect that the top quark is more related to new physics than other fermions 
due to its exceedingly heavy mass. 
Since the top quark nature will be scrutinized at the running Tevatron collider,
the upcoming LHC and the planned LC,  the new physics related to the
top quark sector will be uncovered or constrained.   

One striking property of the top quark in the SM is its superweak FCNC interactions
due to the GIM mechanism: they are absent at tree-level and are extremely
suppressed at loop-level since such FCNC interactions are induced by the charge-current 
CKM transitions involving down-type quarks in the loops which are much lighter than the
top quark. All the top quark FCNC processes in the SM are suppressed to be far below the 
observable level at the existing and upcoming high energy colliders.    
Therefore, the detection of any  top quark FCNC process would serve as a robust 
evidence of new physics beyond the SM. 

In the extensions of the SM, the GIM mechanism usually does not work so well as in
the SM. Some models even predict top quark FCNC Yukawa interactions at tree-level. 
Some top quark FCNC processes may be enhanced to the accessible level at 
colliders.  Searching for these FCNC processes would be a powerful probe for
new physics. 

There are numerous speculations on the possible forms of new physics. Among them 
there are two typical frameworks: one is weak-scale SUSY
which maintains the fundamental scalars while stabilizes the gauge hierarchy; 
the other is technicolor which breaks the electroweak symmetry dynamically.
The most extensively studied SUSY model is the minimal supersymmetric model (MSSM) \Blue{\cite{MSSM}},  
while the most favorable technicolor model is the TC2 model \Blue{\cite{tc2}} which 
combines the idea of topcolor with technicolor.
In this note we briefly review the studies in these two types of new physics models.
The studies in the simplest extension of the SM, i.e., the two-higgs-doublet model
(2HDM), are also listed for comparison.    

\Blue{\section{FCNC top quark processes at a linear collider}}

\Blue{\subsection{FCNC decays}}
Top quark can decay into a charm quark plus a vector boson or 
a light scalar, as shown in Fig.1.  
Almost all known models avoid $t\bar c V$ ($V=\gamma, Z, g$) couplings at tree-level and thus 
$t\to c V$ are induced only at loop-level. 
For the top quark Yukawa interactions, however, FCNC may be exist at tree-level 
in some models like TC2 and 2HDM.   

\begin{center}
\includegraphics[height=5.5cm,width=15cm,angle =0]{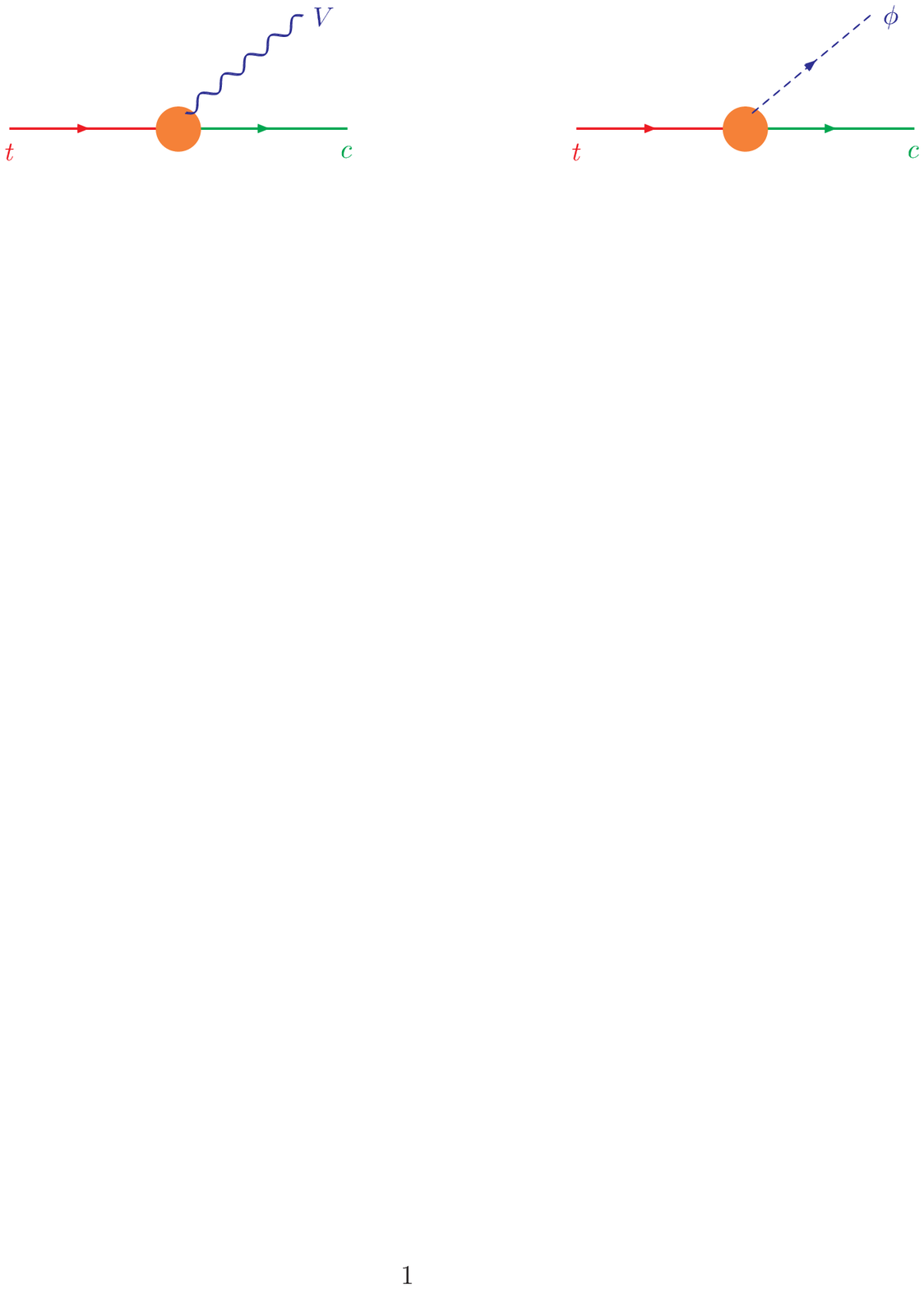}
\vspace*{-1cm} 

{\it Fig.1 Feynman diagrams of FCNC top decays. }
\end{center} 

\Blue{\subsection{Top-charm associated productions}}

Top-charm associated productions at $e^+ e^-$ collision are shown in Fig.2. 
Note $\gamma\gamma \phi$ coupling occurs through loops. The last diagram
may be important in some models since the scalar may be produced on-shell
and the decay $\phi\to t\bar c$ may be dominant.

\begin{center}
\includegraphics[height=17cm,width=15cm,angle =0]{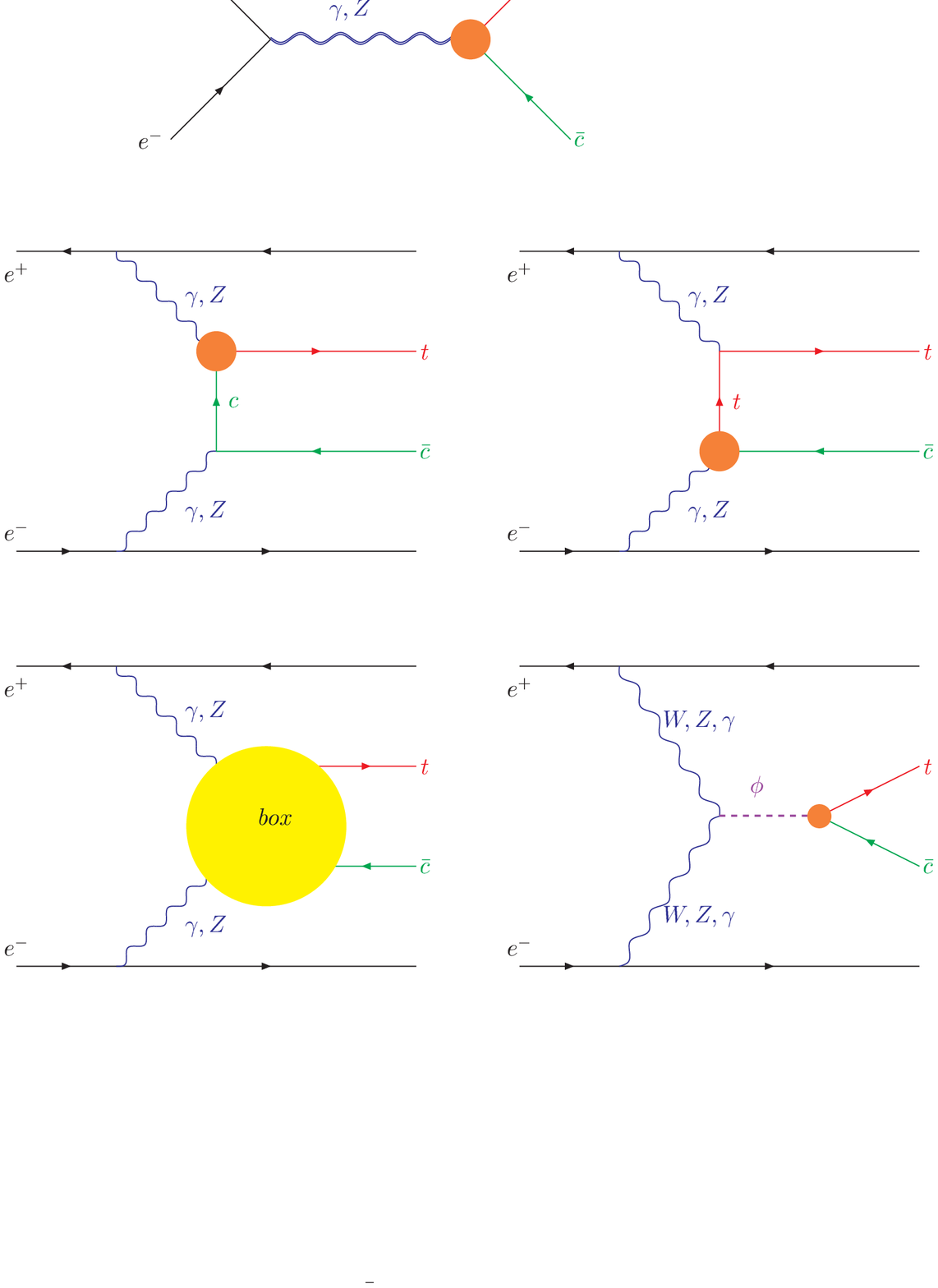}
\vspace*{1cm} 

{\it Fig.2 Feynman diagrams of top-charm associated productions at $e^+ e^-$ collision.} 
\end{center}
\vspace*{1cm}

Top-charm associated productions at $e^-\gamma$ and $\gamma \gamma$ collisions are shown in Figs.3 and 4
respectively. 

\begin{center}
\includegraphics[height=8.5cm,width=15cm,angle =0]{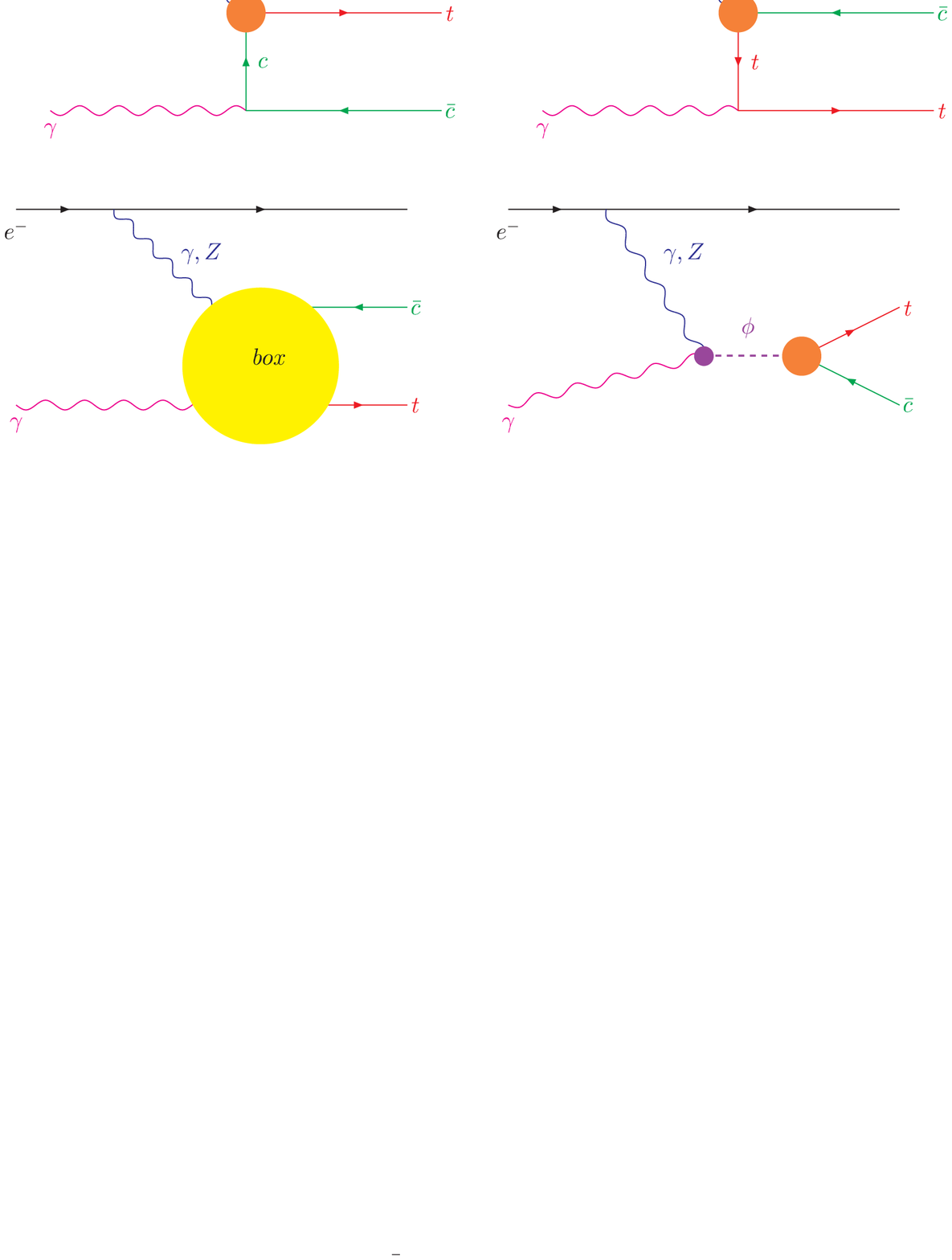}

{\it Fig.3 Feynman diagrams of top-charm associated productions at $e^- \gamma$ collision.} 
\end{center}
\vspace*{0.5cm} 

\begin{center}
\includegraphics[height=8.5cm,width=15cm,angle =0]{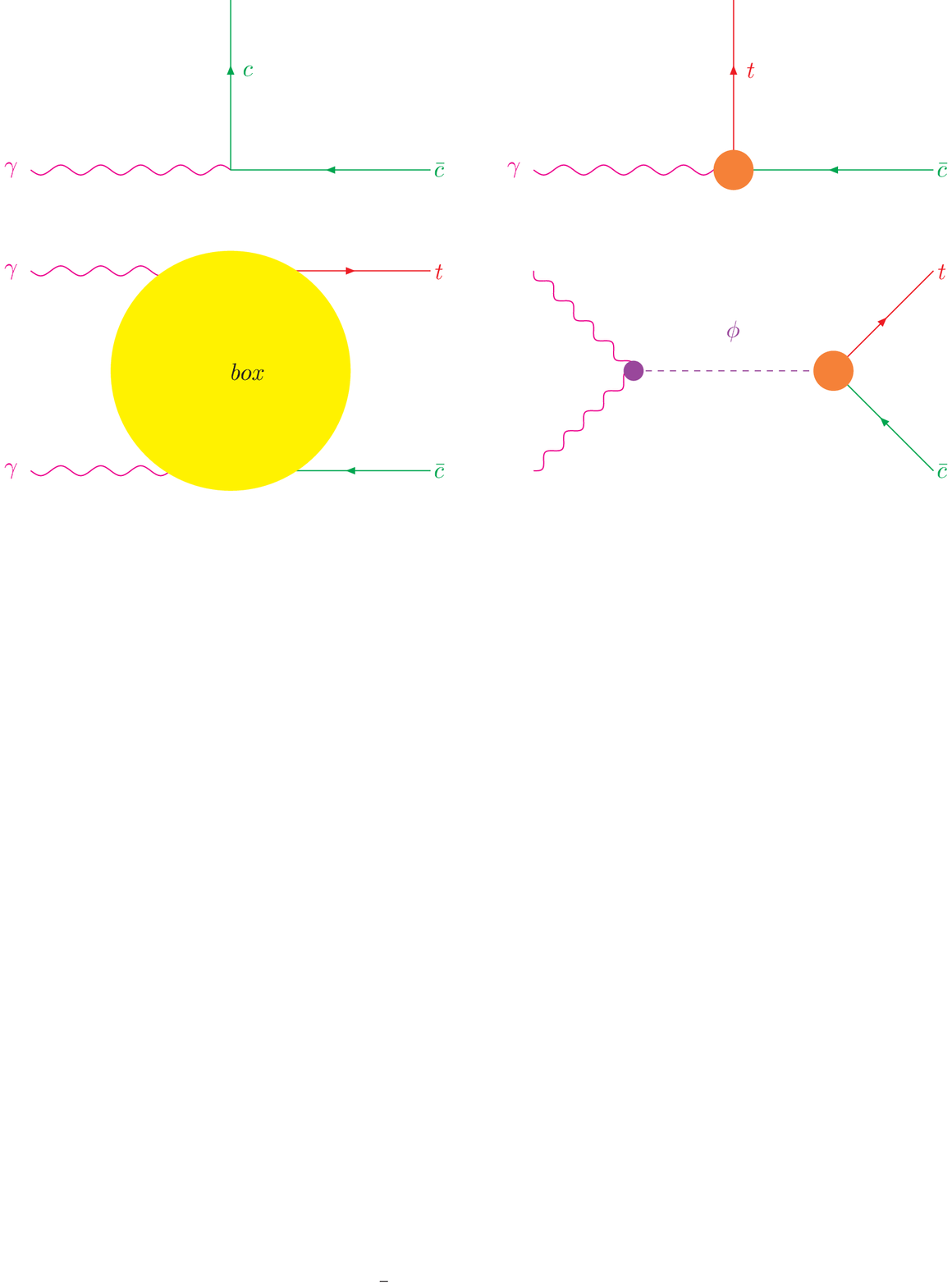}

{\it Fig.4 Feynman diagrams of top-charm associated productions at $\gamma \gamma$ collision.} 
\end{center}

\Blue{\section{Predictions in different models}}
\Blue{\subsection{SM Predictions}}

In the SM the FCNC top-charm transition proceeds through CKM charge-current interactions,
as shown in Fig.5. Such transition is extremely weak since it is proportional to 
$\sim K_{2i} K_{3i}^* f(m_{d_i})$ and $m_{d_i} << m_t$. The top-charm associated 
production rates at a linear collider is shown in Fig.6. 
A summary of the SM predictions is given in Table 1. 
\vspace*{-1cm} 

\begin{center}
\includegraphics[height=5.5cm,width=8cm,angle =0]{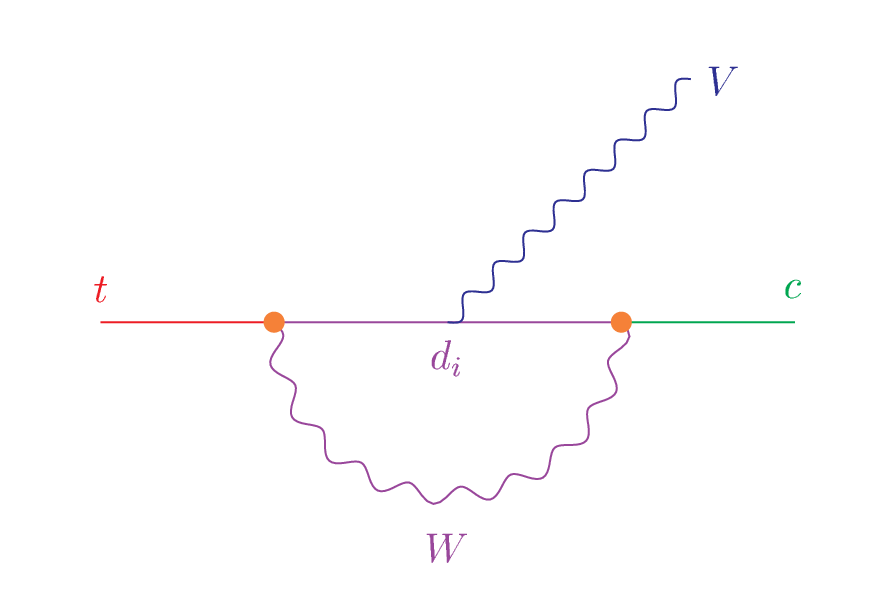}

{\it Fig.5  An example of top-charm transition in the SM.} 
\end{center}

\begin{center}
\includegraphics[height=11cm,width=11cm,angle =0]{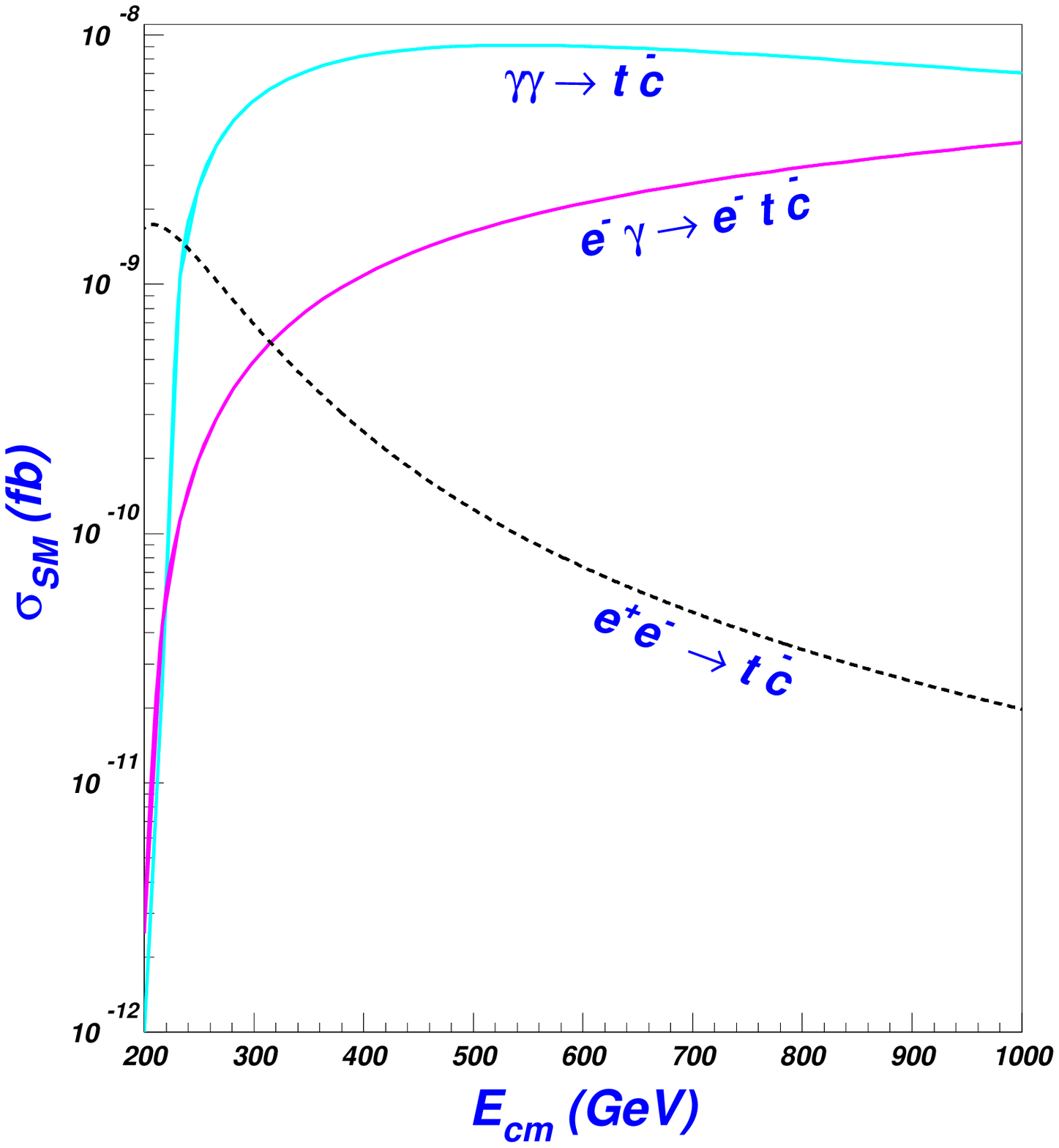}

{\it Fig.6   Top-charm associated production rates at a linear collider in the SM \Blue{\cite{eetc-mssm}}. }
\end{center}

\begin{center}
{\it Table 1: A summary of the SM predictions for FCNC top quark processes.} 
\vspace*{1cm}
\large

\begin{tabular}{ll}
\hline
$ \sigma (\gamma \gamma \to t \bar{c})\ \ \  $ &
   \ \ \  ${\cal{O}}(10^{-8}) $ \Blue{\cite{eetc-mssm}}   \\  \hline
$ \sigma (e \gamma \to e t \bar{c})  $\ \ \ & 
   \ \ \  ${\cal{O}}(10^{-9}) $ \Blue{\cite{eetc-mssm}}   \\ \hline
$ \sigma (e^+ e^- \to t \bar{c}) $\ \ \  &
   \ \ \  ${\cal{O}}(10^{-10})$ \Blue{\cite{sm2,eetc-mssm}} \\ \hline
$ \sigma (e^+ e^- \stackrel{\gamma^\ast \gamma^\ast}{\to} e^+ e^-  t \bar{c}) $\ \ \  &
   \ \ \  $< 10^{-10} $  \Blue{\cite{eetc-mssm}}   \\ \hline
$ \sigma (e^+ e^- \stackrel{Z^\ast Z^\ast}{\to} e^+ e^-  t \bar{c})$\ \ \ & 
   \ \ \  $< 10^{-10}$ \Blue{\cite{cao2}}  \\ \hline
$ \sigma (e^+ e^- \stackrel{\gamma^\ast Z^\ast}{\to} e^+ e^-  t \bar{c}) $ \ \ \ & 
   \ \ \ $ < 10^{-10}$ \Blue{\cite{cao2}}  \\ \hline
$ \sigma (e^+ e^- \stackrel{W^\ast W^\ast}{\to} \nu \bar{\nu}  t \bar{c}) $ \ \ \ & 
   \ \ \ $ < 10^{-10}$ \Blue{\cite{cao2}} \\ \hline
$ Br( t \to c g)$\ \ \  & 
   \ \ \ ${\cal{O}}(10^{-11}) $ \Blue{\cite{sm3}} \\ \hline
$ Br( t \to c Z)  $ \ \ \ & 
   \ \ \ ${\cal{O}}(10^{-13}) $ \Blue{\cite{sm3}} \\ \hline
$ Br( t \to c \gamma) $ \ \ \ & 
   \ \ \ ${\cal{O}}(10^{-13}) $ \Blue{\cite{sm3}} \\ \hline
$ Br( t \to c h) $ \ \ \ & 
   \ \ \ $< 10^{-13} $ \Blue{\cite{sm3}} \\ \hline
\end{tabular}
\end{center}

\Blue{\subsection{SUSY predictions}}

In the MSSM the top-charm FCNC transition can occur through CKM charge-current interaction:
besides the SM-particle loops, there are sparticle loops, as shown in Fig.7 as an example.  
In addition,  the  FCNC transition can be induced by squark flavor mixings, i.e., 
the squark flavor mixings induce the FCNC quark-squark-gluino couplings as well as 
quark-squark-neutralino couplings which in turn induce top-charm transitions as
shown in Fig.8. 

\begin{center}
\includegraphics[height=5.5cm,width=16cm,angle =0]{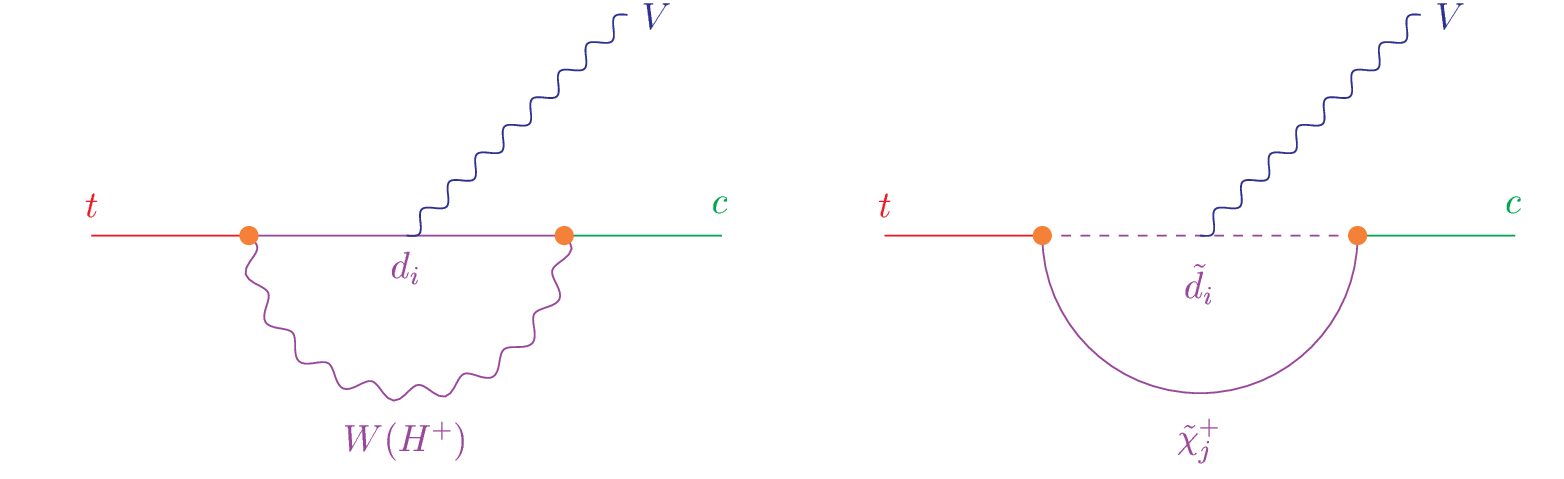}

{\small \it Fig.7  Examples of top-charm transition induced by CKM charge-current in the MSSM.} 
\end{center}

\begin{center}
\includegraphics[height=5.5cm,width=16cm,angle =0]{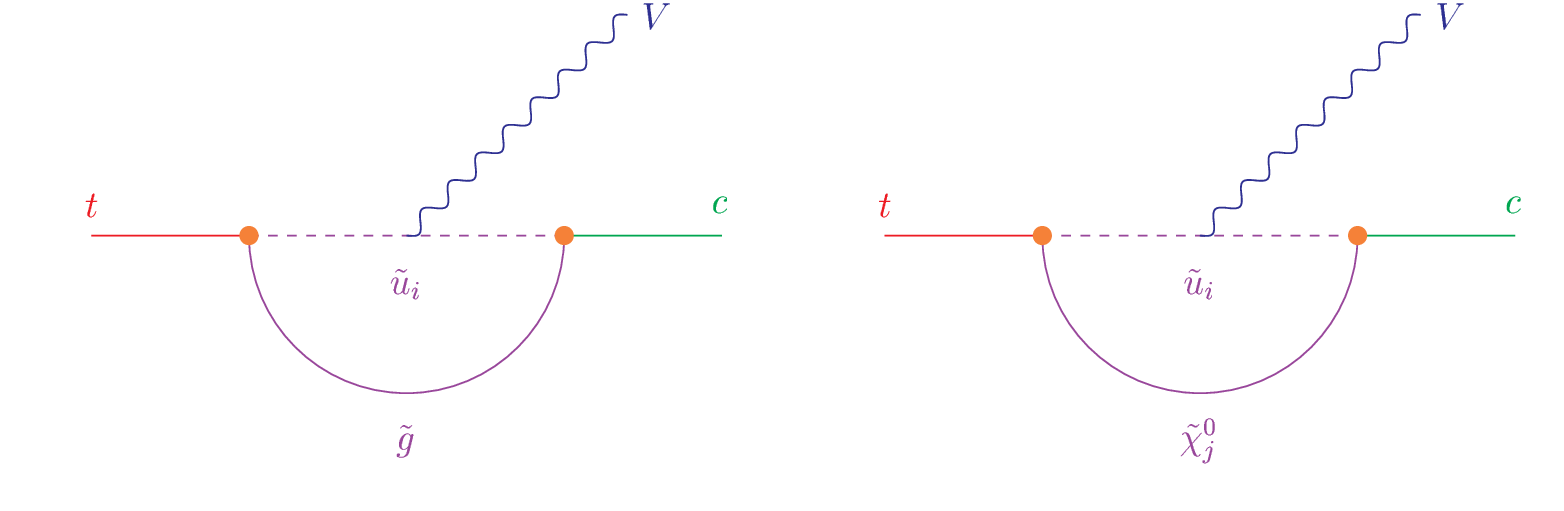}

{\small \it Fig.8  Examples of top-charm transition  induced by FCNC quark-squark-gluino/neutralino in the MSSM.} 
\end{center}

Some scatter plots for the SUSY contributions to top-charm associated production rates
and top rare decays rates at a linear collider with $\sqrt s=500$ GeV are shown in 
Figs.9 and 10. $\delta_L$ is a parameter parameterizing the mixing between  
$\tilde c_L$ and  $\tilde t_L$. A summary of SUSY contributions to FCNC top processes 
is shown in Table 2. 

\begin{center}
\includegraphics[height=11cm,width=5.8cm,angle =0]{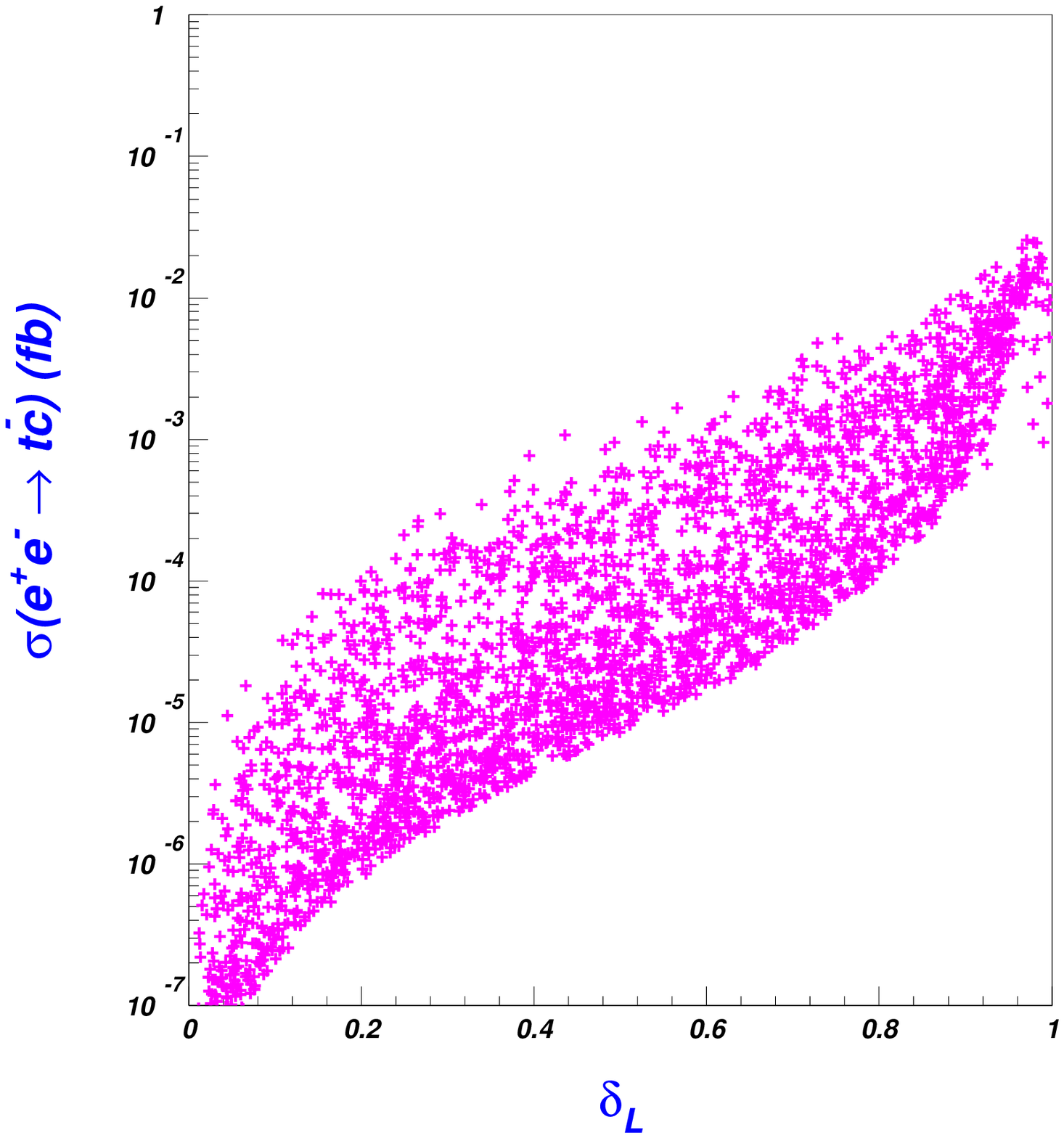}
\includegraphics[height=11cm,width=5.8cm,angle =0]{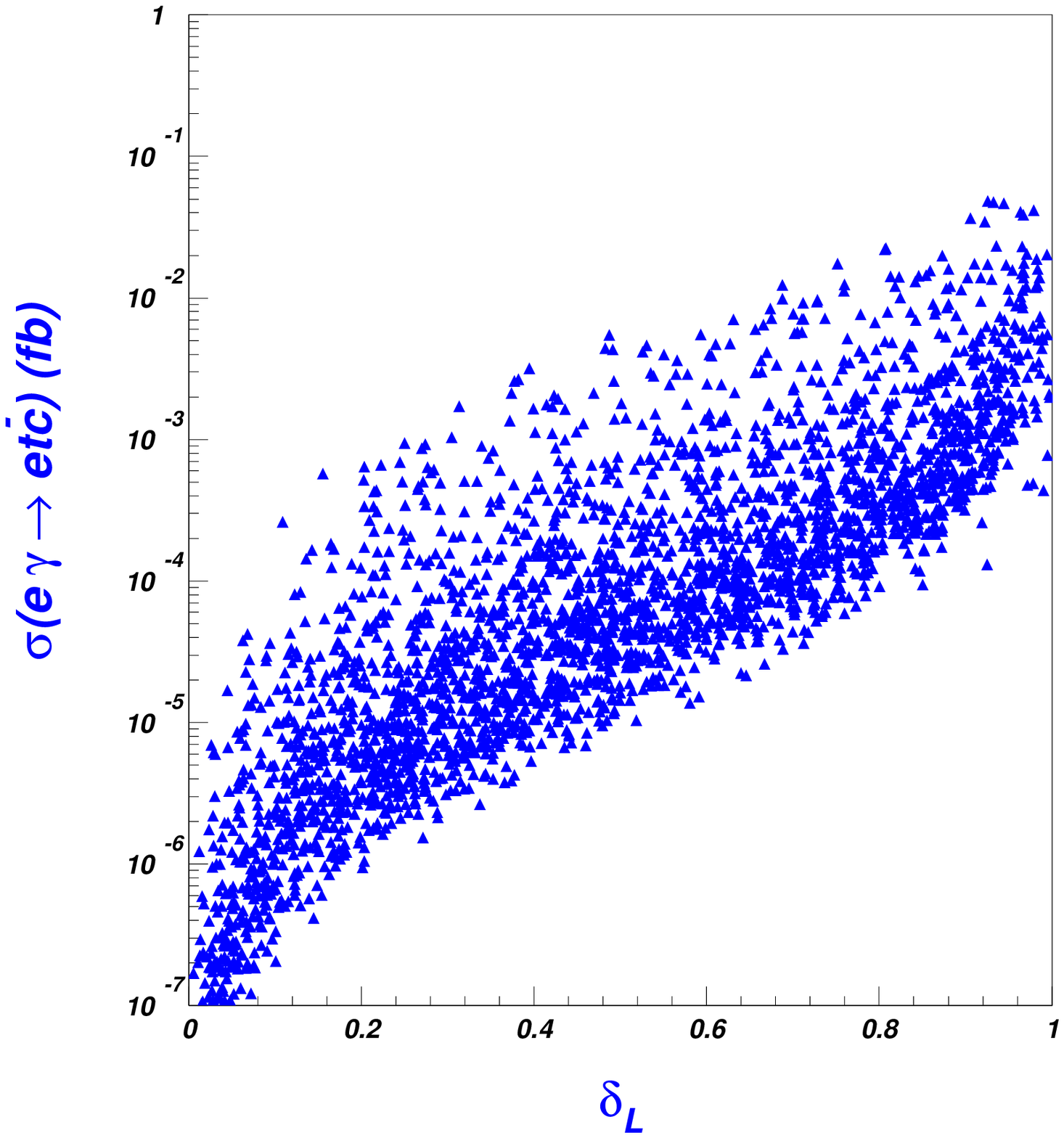}
\includegraphics[height=11cm,width=5.8cm,angle =0]{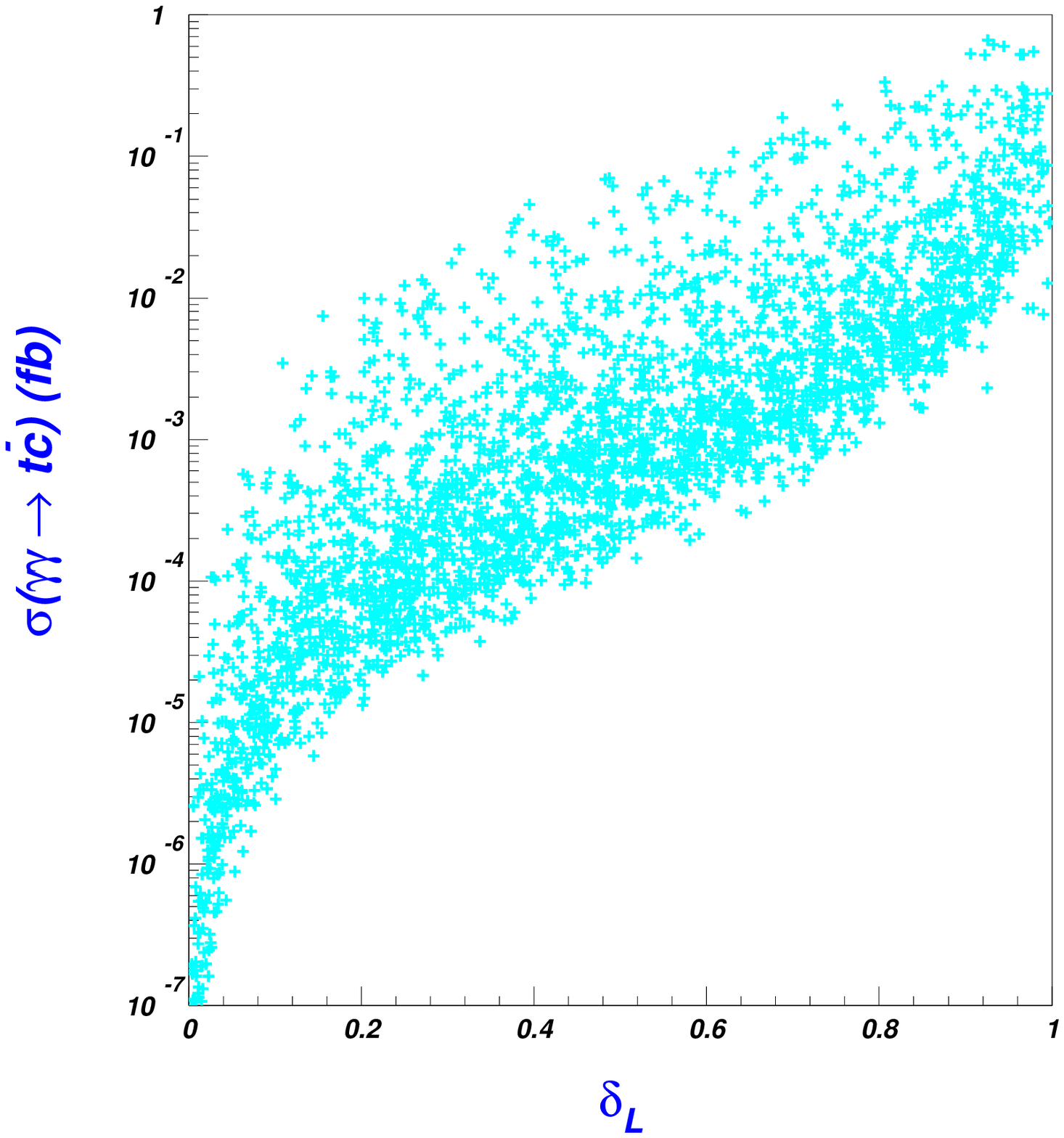}

\small{\it Fig.9 Scatter plots of  SUSY-QCD contribution to $t\bar c$ production rates
           at a linear collider \Blue{\cite{eetc-mssm}}.}
\end{center}

\begin{center}
\includegraphics[height=11cm,width=5.8cm,angle =0]{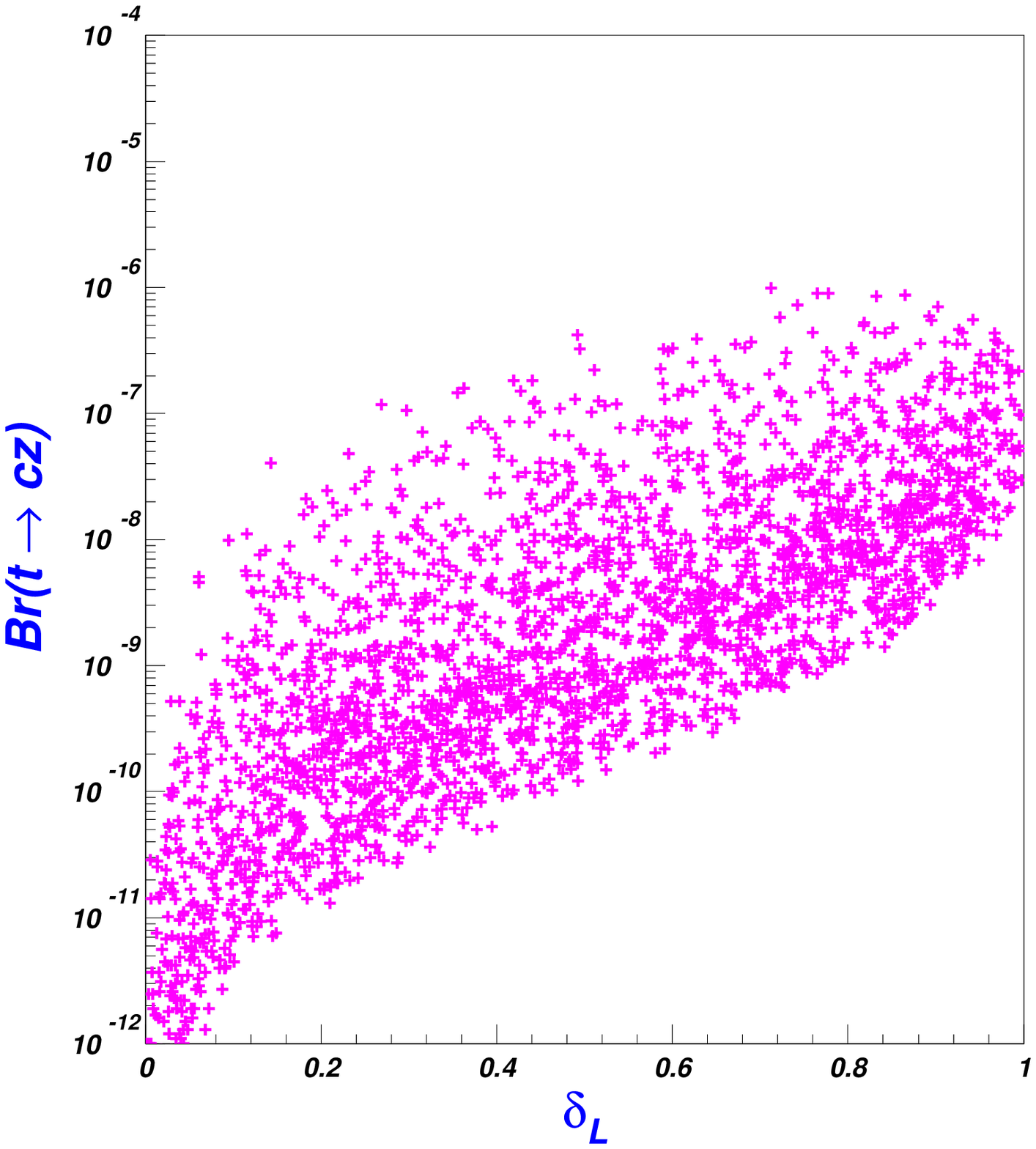}
\includegraphics[height=11cm,width=5.8cm,angle =0]{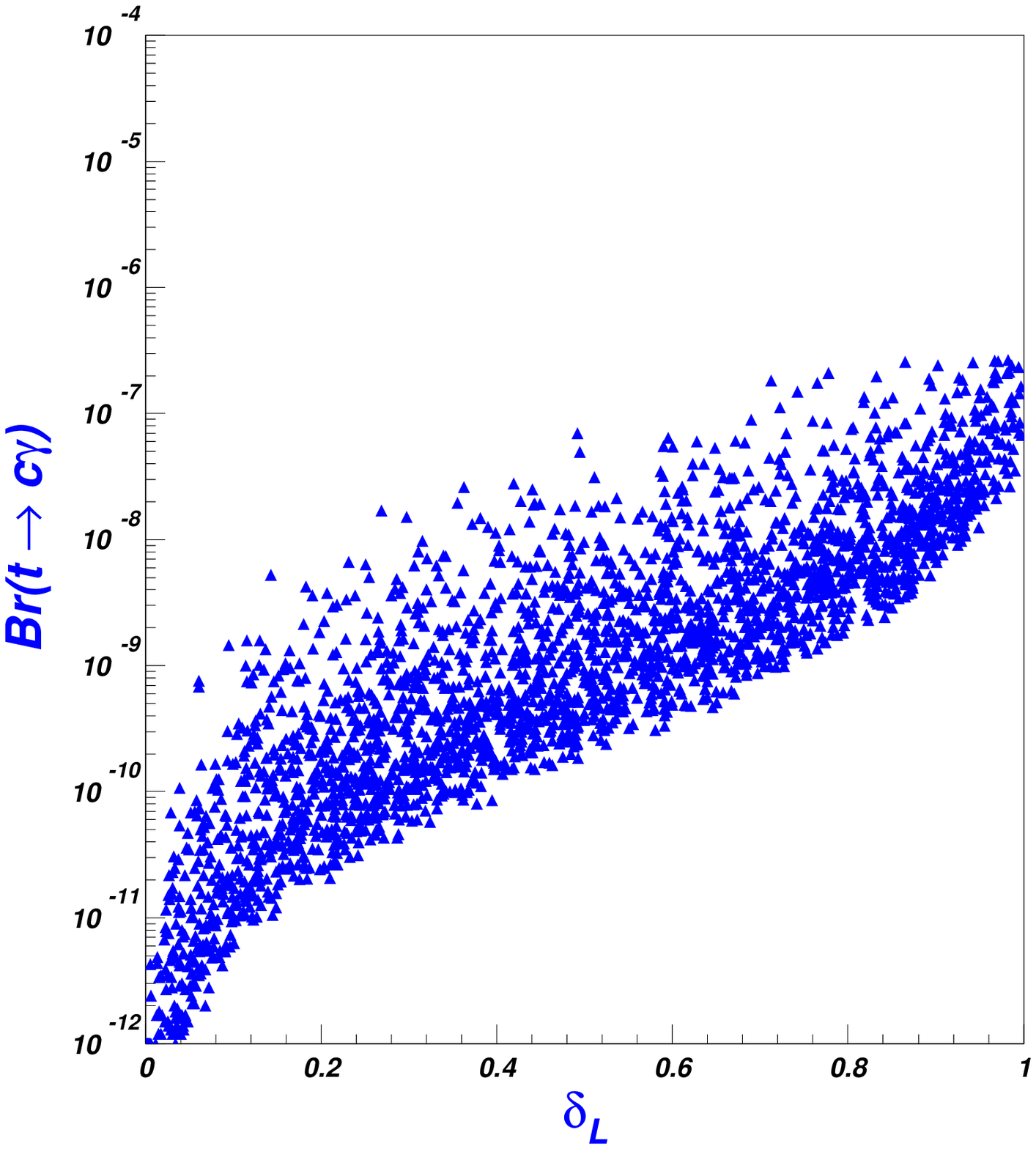}
\includegraphics[height=11cm,width=5.8cm,angle =0]{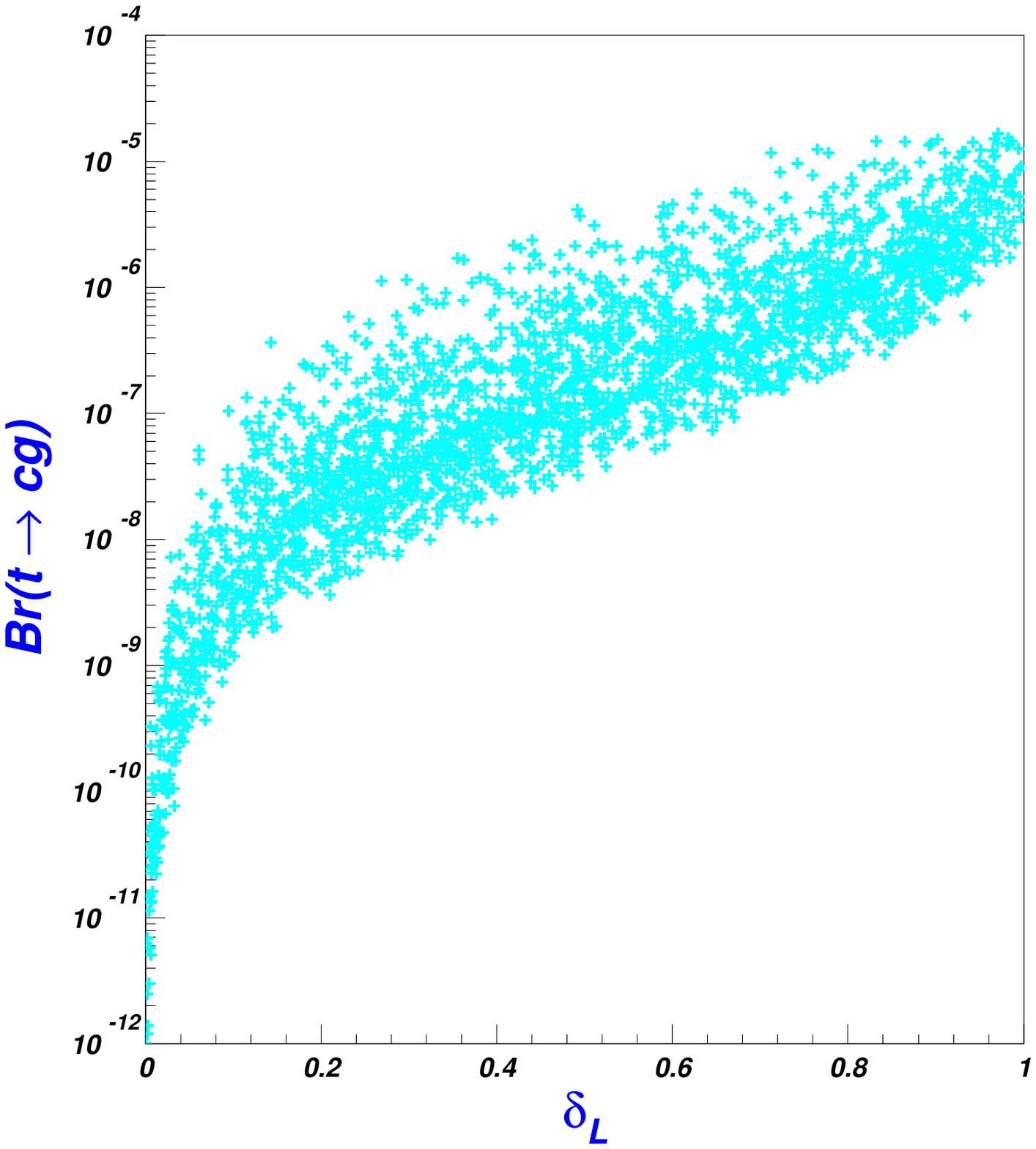}

\small{\it Fig.10 Scatter plots of  SUSY-QCD contribution to FCNC top rare decays \Blue{\cite{eetc-mssm}}. }
\end{center}
\null \noindent \begin{center} {\small \it Table 2: The maximum values of SUSY-QCD predictions 
for top-quark FCNC processes. \\ The electron-positron collider energy 
is $500$ GeV for production processes. }

 \vspace*{0.4cm}
\begin{tabular}{|l|l|l|}
\hline
 &~~~~SM&~~SUSY~~  \\  \hline
$ \sigma (\gamma \gamma \to t \bar{c})\ \ \  $ &
   \ \ \  ${\cal{O}}(10^{-8}) $ \Blue{\cite{eetc-mssm}}\ \ \      & 
   \ \ \  ${\cal{O}}(10^{-1}) $ \Blue{\cite{eetc-mssm}}\ \ \       \\  \hline
$ \sigma (e \gamma \to e t \bar{c})  $\ \ \ & 
   \ \ \  ${\cal{O}}(10^{-9}) $ \Blue{\cite{eetc-mssm}}\ \ \      & 
   \ \ \  ${\cal{O}}(10^{-2}) $ \Blue{\cite{eetc-mssm}}\ \ \      \\ \hline
$ \sigma (e^+ e^- \to t \bar{c}) $\ \ \  &
   \ \ \  ${\cal{O}}(10^{-10})$ \Blue{\cite{sm2,eetc-mssm}} \ \ \ &  
   \ \ \  ${\cal{O}}(10^{-2}) $ \Blue{\cite{eetc-mssm}} \ \ \     \\ \hline
$ \sigma (e^+ e^- \stackrel{\gamma^\ast \gamma^\ast}{\to} e^+ e^-  t \bar{c}) $\ \ \  &
   \ \ \  $< 10^{-10} $  \Blue{\cite{eetc-mssm}} \ \ \            & 
   \ \ \  ${\cal{O}}(10^{-3}) $  \Blue{\cite{eetc-mssm}} \ \ \    \\ \hline
$ \sigma (e^+ e^- \stackrel{Z^\ast Z^\ast}{\to} e^+ e^-  t \bar{c})$\ \ \ & 
   \ \ \  $< 10^{-10}$ \Blue{\cite{cao2}}   \ \ \                 & 
   \ \ \  $ < 10^{-3}$ \Blue{\cite{cao2}}                         \\ \hline
$ \sigma (e^+ e^- \stackrel{\gamma^\ast Z^\ast}{\to} e^+ e^-  t \bar{c}) $ \ \ \ & 
   \ \ \ $ < 10^{-10}$ \Blue{\cite{cao2}} \ \ \                   & 
   \ \ \ $ < 10^{-3} $ \Blue{\cite{cao2}}                         \\ \hline
$ \sigma (e^+ e^- \stackrel{W^\ast W^\ast}{\to} \nu \bar{\nu}  t \bar{c}) $ \ \ \ & 
   \ \ \ $ < 10^{-10}$ \Blue{\cite{cao2}} \ \ \                   & 
   \ \ \ $ < 10^{-3} $ \Blue{\cite{cao2}}                         \\ \hline
$ Br( t \to c g)$\ \ \  & 
   \ \ \ ${\cal{O}}(10^{-11}) $ \Blue{\cite{sm3}} \ \ \           & 
   \ \ \ ${\cal{O}}(10^{-5}) $ \Blue{\cite{tcv,eetc-mssm}}\ \ \   \\ \hline
$ Br( t \to c Z)  $ \ \ \ & 
   \ \ \ ${\cal{O}}(10^{-13}) $ \Blue{\cite{sm3}}\ \ \            & 
   \ \ \ ${\cal{O}}(10^{-7}) $ \Blue{\cite{tcv,eetc-mssm}}\ \ \   \\ \hline
$ Br( t \to c \gamma) $ \ \ \ & 
   \ \ \ ${\cal{O}}(10^{-13}) $ \Blue{\cite{sm3}} \ \ \           &
   \ \ \ ${\cal{O}}(10^{-7}) $ \Blue{\cite{tcv,eetc-mssm}}\ \ \   \\ \hline
$ Br( t \to c h) $ \ \ \ & 
   \ \ \ $< 10^{-13} $ \Blue{\cite{sm3}} \ \ \                    & 
   \ \ \ ${\cal{O}}(10^{-4})$ \Blue{\cite{MSSM3}}\ \ \            \\ \hline
\end{tabular}
\end{center}

\Blue{\subsection{Technicolor  (TC2) predictions }}

The top quark FCNC may be severe in TC2 model for the following reason:
\begin{itemize}
\item Topcolor is non-universial, only cause top-quark to condensate, only give top-quark mass 
      (a large portion $1-\epsilon_t$); neutral top-pions have large Yukawa couplings to only top quark; 
\item ETC gives masses to all quarks (for top quark, only a small portion $\epsilon_t$); 
      ETC-pions have small Yukawa couplings to all quarks 
      (for top-quark, coupling is much weaker than top-pion's);  
\item So the mass matrix of up-type quarks is composed of both ETC and Topcolor contributions.
      We use unitary matrices $U_L$ and $U_R$ to rotate left- and right-handed up-type quarks 
      in order to diagonalize this mass matrix.  
\item Clearly such $U_L$ and $U_R$ rotations will make top-quark top-pion Yukawa couplings 
      ( $i\pi^0_t \bar t_L t_R$) to have FCNC.          

 (Of course, if ETC does not give mass to top-quark ($\epsilon_t=0$), then no FCNC for top-quark).   
\end{itemize}

Top-charm transition can then occur through top-pion Yukawa couplings, as shown in Fig.11. 
$t\bar c$ production rate in TC2 is shown in Fig.12.
A summary of predictions of different models for FCNC top-quark processes are
shown in Table 3. The predictions in the general two-Higgs doublet model,
 the so-called "type-III" model (2HDM-III), are also listed.
In the 2HDM-III the up-type and down-type quarks couple to both Higgs doublets and 
thus the FCNC usually exists in the Yukawa couplings. 
\vspace*{1.5cm}

\begin{center}
\includegraphics[height=5.5cm,width=16cm,angle =0]{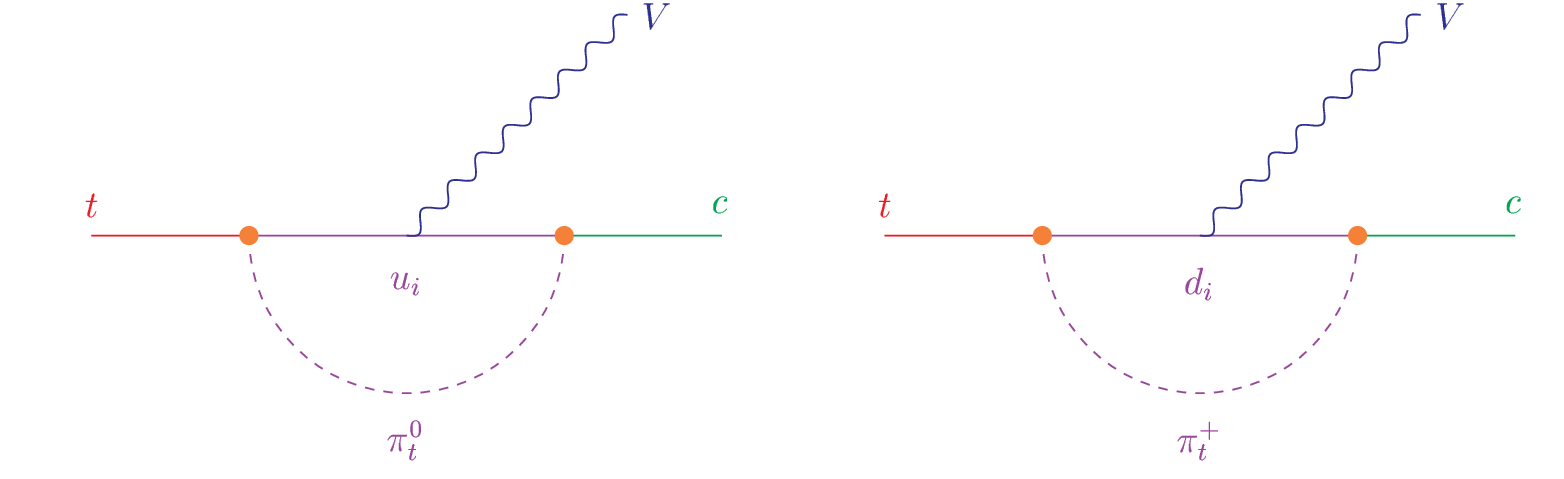}

{\it Fig.11  Examples of top-charm transition  induced by top-pion Yukawa couplings in TC2 model.} 
\end{center}

\begin{center}
\includegraphics[height=7.5cm,width=7.5cm,angle =0]{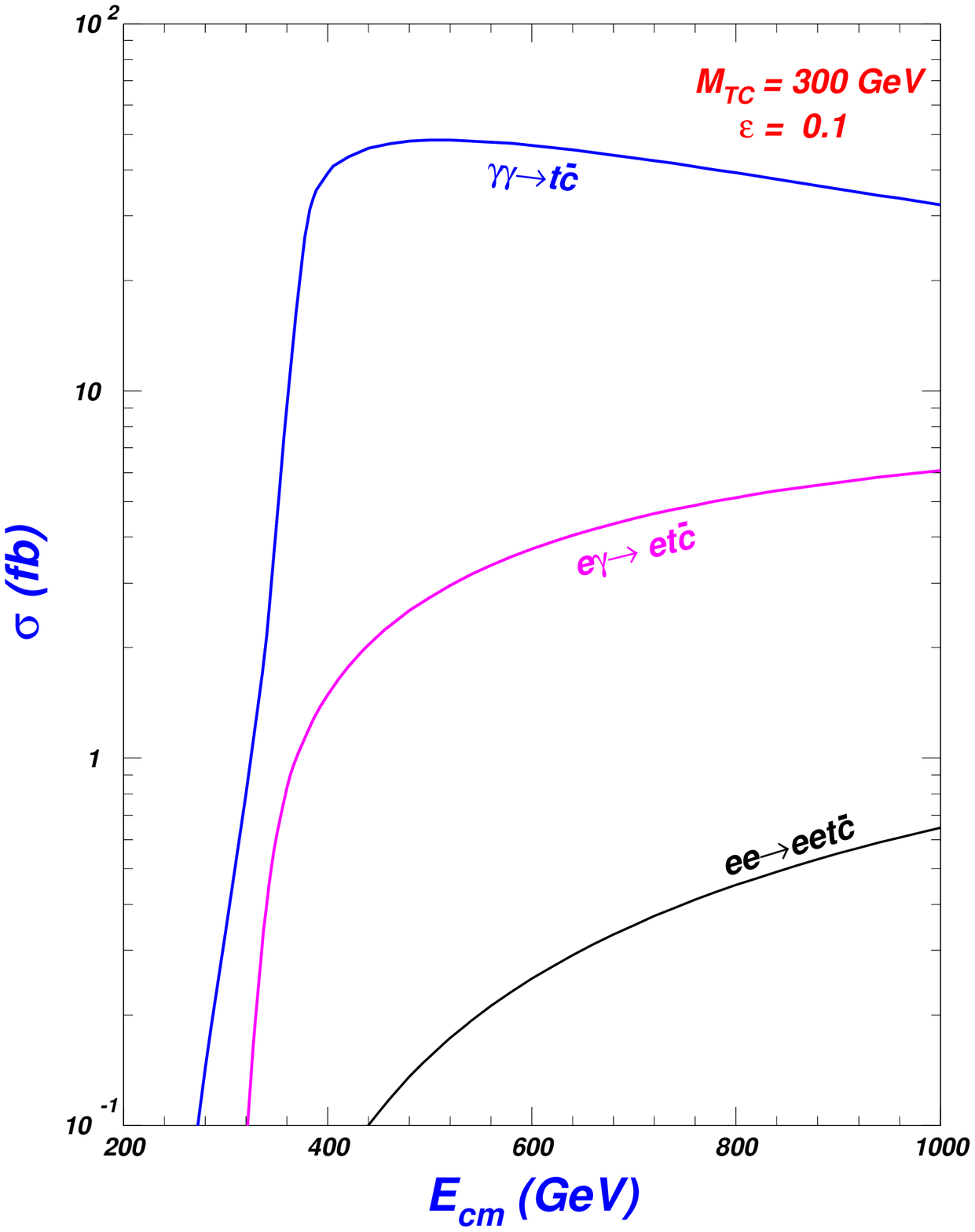}
\includegraphics[height=7.5cm,width=7.5cm,angle =0]{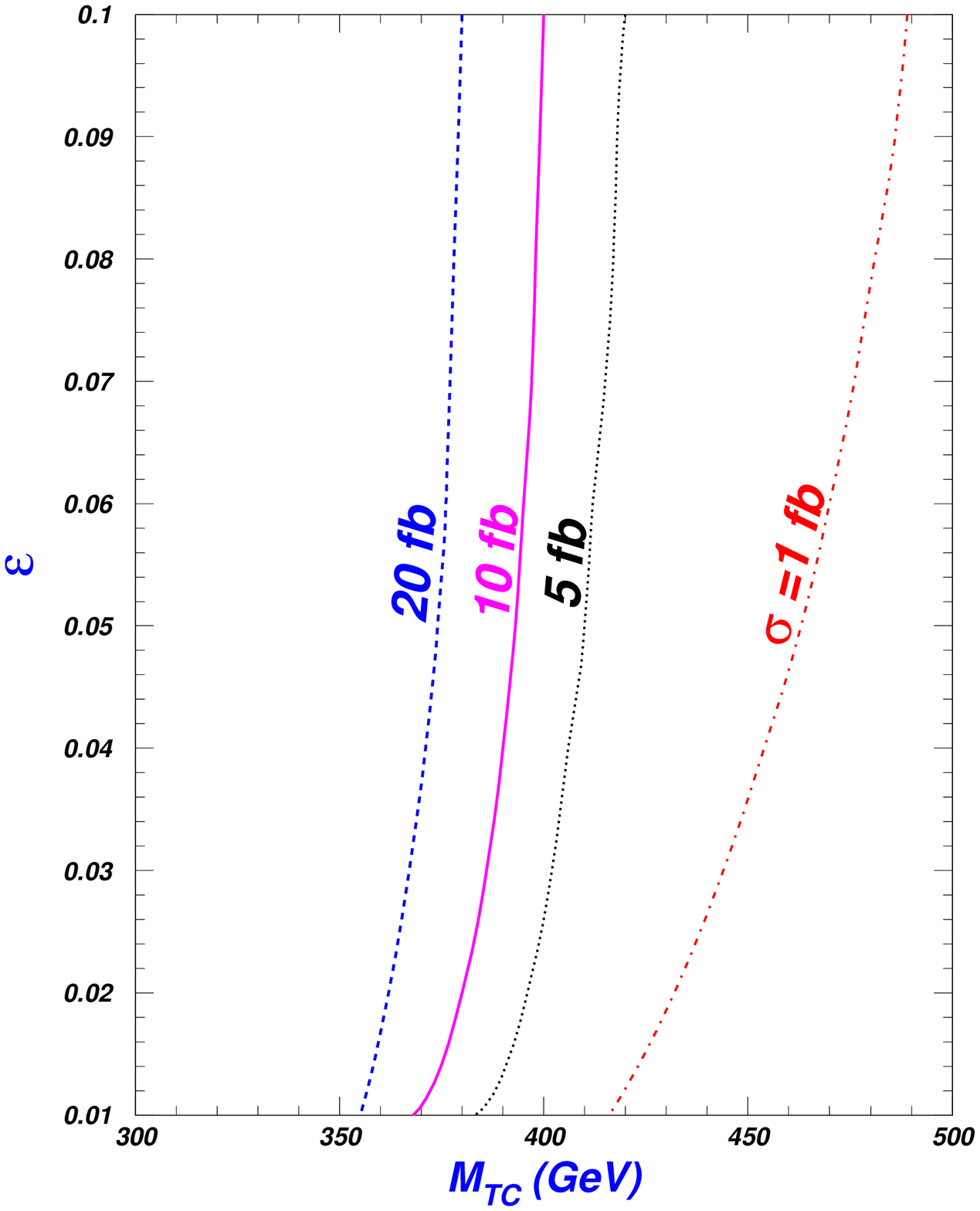}

{\it Fig.12   $t \bar c$ production rate at a linear $e^+e^-$ collider with 
                     $\sqrt{s}=500$ GeV in TC2 model \Blue{\cite{cao2}}.} 
\end{center}
 
\null \noindent {\small Table 3: Maximum predictions for
top-quark FCNC processes.  The $e^+e^-$ collider energy is 500 GeV for
production processes. The cross sections are in the units of fb.} \vspace*{-0.3cm}
\begin{center}
\begin{tabular}{lllll}
\hline
& ~~~SM  & ~~~2HDM-III & ~~~MSSM & ~~~TC2 \\
 \hline 
$ \sigma (\gamma \gamma \to t \bar{c})\ \ \  $ &
   \ \ \  ${\cal{O}}(10^{-8}) $ \Blue{\cite{eetc-mssm}}\ \ \      & \ \ \  ${\cal{O}}(10^{-1}) $\Blue{\cite{ma2}}\ \ \    & 
   \ \ \  ${\cal{O}}(10^{-1}) $ \Blue{\cite{eetc-mssm}}\ \ \      & \ \ \  ${\cal{O}}(10)$ \Blue{\cite{cao2}} \ \ \ \\  \hline
$ \sigma (e \gamma \to e t \bar{c})  $\ \ \ & 
   \ \ \  ${\cal{O}}(10^{-9}) $ \Blue{\cite{eetc-mssm}}\ \ \      & \ \ \  ${\cal{O}}(10^{-2})$ \Blue{\cite{cao2}} \ \ \    &
   \ \ \  ${\cal{O}}(10^{-2}) $ \Blue{\cite{eetc-mssm}}\ \ \      & \ \ \  ${\cal{O}}(1)$  \Blue{\cite{cao2}} \ \ \ \\ \hline
$ \sigma (e^+ e^- \to t \bar{c}) $\ \ \  &
   \ \ \  ${\cal{O}}(10^{-10})$ \Blue{\cite{sm2,eetc-mssm}} \ \ \ &  \ \ \  ${\cal{O}}(10^{-3}) $ \Blue{\cite{2hdm2}}\ \ \   & 
   \ \ \  ${\cal{O}}(10^{-2}) $ \Blue{\cite{eetc-mssm}} \ \ \     & \  \ \ ${\cal{O}}(10^{-1}) $  \Blue{\cite{eetc-tc2}}\ \ \ \\ \hline
$ \sigma (e^+ e^- \stackrel{\gamma^\ast \gamma^\ast}{\to} e^+ e^-  t \bar{c}) $\ \ \  &
   \ \ \  $< 10^{-10} $  \Blue{\cite{eetc-mssm}} \ \ \            & \ \ \  ${\cal{O}}(10^{-3})$ \Blue{\cite{cao2}}   \ \ \ & 
   \ \ \  ${\cal{O}}(10^{-3}) $  \Blue{\cite{eetc-mssm}} \ \ \    & \ \ \  ${\cal{O}}(10^{-1})$ \Blue{\cite{cao2}}   \ \ \  \\ \hline
$ \sigma (e^+ e^- \stackrel{Z^\ast Z^\ast}{\to} e^+ e^-  t \bar{c})$\ \ \ & 
   \ \ \  $< 10^{-10}$ \Blue{\cite{cao2}}   \ \ \                 & \ \ \  ${\cal{O}}(10^{-1}) $  \Blue{\cite{2hdm3}}  \ \ \   & 
   \ \ \  $ < 10^{-3}$ \Blue{\cite{cao2}}                         & \ \ \  ${\cal{O}}(1)$\Blue{\cite{ee-ht}} \ \ \ \\ \hline
$ \sigma (e^+ e^- \stackrel{\gamma^\ast Z^\ast}{\to} e^+ e^-  t \bar{c}) $ \ \ \ & 
   \ \ \ $ < 10^{-10}$ \Blue{\cite{cao2}} \ \ \                   & \ \ \  $ < 10^{-3}$ \Blue{\cite{cao2}} \ \ \   & 
   \ \ \ $ < 10^{-3} $ \Blue{\cite{cao2}}                         & \  \ \ $<10^{-1} $\Blue{\cite{cao2}}  \ \ \ \\ \hline
$ \sigma (e^+ e^- \stackrel{W^\ast W^\ast}{\to} \nu \bar{\nu}  t \bar{c}) $ \ \ \ & 
   \ \ \ $ < 10^{-10}$ \Blue{\cite{cao2}} \ \ \                   & \ \ \  ${\cal{O}}(10^{-1}) $  \Blue{\cite{2hdm3}}  \ \ \ &  
   \ \ \ $ < 10^{-3} $ \Blue{\cite{cao2}}                         & \ \ \ ${\cal{O}}(1) $  \Blue{\cite{ee-ht}} \ \ \ \\ \hline
$ Br( t \to c g)$\ \ \  & 
   \ \ \ ${\cal{O}}(10^{-11}) $ \Blue{\cite{sm3}} \ \ \           & \ \ \ ${\cal{O}}(10^{-5}) $ \Blue{\cite{2hdm4}}\ \ \ &
   \ \ \ ${\cal{O}}(10^{-5}) $ \Blue{\cite{tcv,eetc-mssm}}\ \ \   & \ \ \ ${\cal{O}}(10^{-4}) $ \Blue{\cite{tctcv}}\ \ \  \\ \hline
$ Br( t \to c Z)  $ \ \ \ & 
   \ \ \ ${\cal{O}}(10^{-13}) $ \Blue{\cite{sm3}}\ \ \            & \ \ \ ${\cal{O}}(10^{-6}) $ \Blue{\cite{2hdm4}} \ \ \ &
   \ \ \ ${\cal{O}}(10^{-7}) $ \Blue{\cite{tcv,eetc-mssm}}\ \ \   & \ \ \ ${\cal{O}}(10^{-4}) $ \Blue{\cite{tctcv}}\ \ \ \\ \hline
$ Br( t \to c \gamma) $ \ \ \ & 
   \ \ \ ${\cal{O}}(10^{-13}) $ \Blue{\cite{sm3}} \ \ \           & \ \ \ ${\cal{O}}(10^{-7}) $ \Blue{\cite{2hdm4}} \ \ \ &  
   \ \ \ ${\cal{O}}(10^{-7}) $ \Blue{\cite{tcv,eetc-mssm}}\ \ \   & \ \ \ ${\cal{O}}(10^{-6}) $ \Blue{\cite{tctcv}}\ \ \ \\ \hline
$ Br( t \to c h) $ \ \ \ & 
   \ \ \ $< 10^{-13} $ \Blue{\cite{sm3}} \ \ \                    & \ \ \ ${\cal{O}}(10^{-3}) $ \Blue{\cite{cao2}} \ \ \ &  
   \ \ \ ${\cal{O}}(10^{-4})$ \Blue{\cite{MSSM3}}\ \ \            & \ \ \ ${\cal{O}}(10^{-1}) $ \Blue{\cite{cao2}} \ \ \ \\ \hline
\end{tabular}
\end{center}

\Blue{\section{Conclusion}}

Our conclusions are: 
\begin{itemize}
\item[(1)] The SM predictions for the FCNC top quark processes at a linear collider are far 
           below the observable level.
\item[(2)] The new physics models (like SUSY, TC2) can enhance the SM rates of the FCNC top quark processes
           by several orders.         
\item[(3)] $ \sigma (\gamma \gamma \to t \bar{c}) \Red{>} 
           \sigma (e \gamma \to e t \bar{c})  \Red{>}  \sigma (e^+ e^- \to t \bar{c}) $.
\item[(4)] TC2 predictions \Red{$>>$}  SUSY predictions \Red{$>>$} SM predictions.
\item[(5)] TC2 predictions may be accesible at a linear collider.  
\end{itemize}

Note that we did not discuss the following issues which may be relevant:
\begin{itemize}
\item The sensitivity of an LC to top quark FCNC processes \Blue{\cite{sensitivity}}.
\item Top quark FCNC decays in other new physics models \Blue{\cite{tcvh-other}}.
\item CP violating effects in top quark FCNC productions at an LC \Blue{\cite{cpv-fcnc}}.
\item SUSY or TC2 induced top quark rare processes at hadron collider \Blue{\cite{new-production-pp}}
      and $e p$ collider \Blue{\cite{new-production-ep}}.
\end{itemize}

\end{document}